\documentclass[a4paper,11pt]{article}
\usepackage{graphicx}
\usepackage{epsfig}
\usepackage{subfig}
\usepackage{float}
\usepackage{amsmath}
\usepackage{amsfonts,amssymb}
\usepackage{amssymb}
\usepackage{latexsym}
\usepackage{url}
\usepackage{natbib}
 
\begin{document}

\newtheorem{theorem}{Theorem}[section]
\newtheorem{corollary}[theorem]{Corollary}
\newtheorem{lemma}[theorem]{Lemma}
\newtheorem{proposition}[theorem]{Proposition}
\newtheorem{remark}[theorem]{Remark}
\newtheorem{conjecture}[theorem]{Conjecture}
\newtheorem{example}[theorem]{Example}
\newcommand{\ind}{1\hspace{-2.7mm}{1}}
\newcommand{\parrow}{\xrightarrow[n \to \infty]{p_{\nu}}}
\newcommand{\asarrow}{\xrightarrow{a.s.}}
\newcommand{\Poisson}{\mathcal{P}}
\title{Stochastic SIR epidemics in a population with households and schools \thanks{PT is supported by Vetenskapsr{\aa}det (Swedish Research Council) Grant nr: 2010587
 }}
\author{Tanneke Ouboter, Ronald Meester and Pieter Trapman}

\maketitle

\begin{abstract}
We study the spread of stochastic SIR (Susceptible $\to$ Infectious $\to$ Recovered) epidemics in two types of structured populations, both consisting of schools and households. In each of the types, every individual is part of one school and one household. In the {\em independent partition model}, the partitions of the population into schools and households are independent of each other. This model corresponds to the well-studied household-workplace model. In the {\em hierarchical model} which we introduce here, members of the same household are also members of the same school. 

We introduce computable branching process approximations for both types of populations and use these to compare the probabilities of a large outbreak. The branching process approximation in the hierarchical model is novel and of independent interest. We prove by a coupling argument that if all households and schools have the same size, an epidemic spreads easier (in the sense that the number of individuals infected is stochastically larger) in the independent partition model. We also show by example that this result does not necessarily hold if households and/or schools do not all have the same size. 
\end{abstract}

\section{Introduction and context}

Mathematical modeling of the spread of infectious diseases has a long history \citep{Diek12}. A commonly used model for epidemics, the SIR epidemic in a closed population, is easy to describe, but this model already has interesting features. In stochastic models for epidemics in large, unstructured homogeneously mixing populations (that is, every pair of individuals makes contacts at the same rate) branching process approximations can be used to compute the probability that an epidemic will occur if a disease is introduced in a population. In case of an epidemic, we can also use branching process approximations to compute the expected fraction of the population that is infected throughout the course of the epidemic. 

The homogeneous mixing assumption is too strong. One way to gain realism is to assume multiple levels of mixing. A popular extension is the so-called household model \citep{Ball97}. In this model the population is partitioned into households of relatively small size. Within households, contacts are more frequent than in the general population. Inspired by the modeling of the spread of childhood diseases, this model has been extended further by also introducing an independent partition of the population into schools (or workplaces) \citep{Ball02,Pell09,Pell12}. For these diseases, the spread in schools plays an important role and the household-school model is a natural model in this case.

The assumption that the partition in households and schools are independent has as a consequence that in large populations it is unlikely that members from the same family attend the same school. However, it would be more realistic to assume that siblings do go to the same school. This idea leads to the hierarchical model which we define below. 

Mathematically speaking, the independent partition model is the easier one to understand. It would be of great theoretical and practical interest if we could show that epidemics spread more easily (the precise meaning of this is explained in the following sections) in the independent partition model than in the more realistic hierarchical model. Indeed, control strategies which are known to work in the independent model, will then also stop epidemics in the hierarchical model.

We show in Theorem \ref{mainthm} that in case the sizes of schools are all the same and the sizes of households are all the same, then indeed an epidemic spreads easier in the independent partition model. We use branching process approximations for this conclusion.  However, if the sizes of schools and households are variable, then, perhaps surprisingly, this is not true in general, see Theorem \ref{mainthm2}. The branching process approxiamtion in the hierarchical model is new and interesting in its own right.

\section{The model and  main results}\label{modelsec}

\subsection{Social structure}

We consider two models for a population structure. In either model, individuals are part of exactly one household and exactly one school. In the {\em independent partition model}, the partitions of the population into households and into schools are independent. This model has been studied as the household-workplace model before in e.g.\ \citep{Ball02,Pell09,Pell12}. In the {\em hierarchical model}, members of the same household attend the same school.

We can formally construct populations of either type with a given number of $n$ schools as follows. In the hierarchical model every school contains individuals from an independent and identically distributed (i.i.d.) number of households, where this number of households the individuals in a school are part of is distributed as $N_c$. Households have i.i.d.\ sizes, distributed as $N_h$. Hence the number of individuals in a school is distributed as $N_{s} \sim \sum_{k=1}^{N_c} N_h^{(k)}$, where $N_h^{(k)}$, $k=1,2,\ldots$ are independent copies of $N_h$, which are also independent of $N_c$.
For mathematical convenience we assume that both $N_h$ and $N_c$ have bounded support on the positive integers.  We denote by $N=N(n)$ the total number of individuals in the population, that is, $N = \sum_{k=1}^n N^{(k)}_{s}$, where $N^{(k)}_{s}$ denotes the number of individuals  in the $k$-th school.

In the independent partition model we use the school and household sizes from the hierarchical model and use the independent partitions of the $N$ individuals in the population, uniformly chosen among all partitions with respectively the required household sizes and required school sizes.

\subsection{SIR epidemics}

We consider a stochastic SIR epidemic in a closed population. In this model, individuals are in one of the three states, $S$, $I$ and $R$. If a susceptible individual contacts (note Remark \ref{contdis} below) an infectious individual, the susceptible one becomes infectious immediately (she is \emph{infected}) and stays so for exactly one time unit (but note Remark \ref{Kuuldis} below). After this infectious period the individual recovers and stays immune forever. There are three types of contacts: pairs of individuals which are in the same household make household-contacts according to Poisson processes with intensity $\lambda_h$. Similarly, pairs of individuals within the same school make school-contacts according to Poisson processes with intensity $\lambda_s$. Finally, all pairs of individuals in the population make global-contacts according to Poisson processes with intensity $\lambda_g/(N-1)$. All Poisson processes are independent of each other. Individuals which are in the same household and in the same school have a total contact intensity of $\lambda_h + \lambda_s + \lambda_g/(N-1)$. We assume that initially there is a randomly chosen individual with a remaining infectious period of 1 time unit and all other individuals are susceptible. We assume that the spread of the epidemic - if it spreads - is so fast, that ``demographic'' processes such as births, individuals reaching the school-going age or moving to another school during the epidemic can be ignored.

A \emph{household epidemic} is defined as an outbreak which occurs if all global and school contacts are ignored; a household epidemic is always restricted to one household. Similarly a \emph{school epidemic} is an outbreak which occurs if all global and household contacts are ignored.

In this paper we are mainly interested in the \emph{final size} of an epidemic, that is, the fraction $\rho$ of individuals  which are infected throughout the epidemic. We use (and state the arguments for this) that in the large population limit, if the fraction of infected individuals is positive (i.e.\ a \emph{major outbreak} occurs) than with probability tending to 1 (as the population size grows) this fraction is equal to the  probability of a major outbreak.

\begin{remark}\label{contdis} {\rm Contacts as defined above, are not necessarily identical to physical contacts. Only encounters which lead to the transmission of the disease if one of the individuals is infectious and the other susceptible are considered to be contacts. So, if only half of the contacts of an infectious individual with a susceptible leads to transmission, then we can model the ``infectious contacts'' by thinning all original Poisson processes representing physical encounters. The remaining points are still distributed according to a Poisson process, now with half the density of the original process.}
\end{remark}
\begin{remark}\label{Kuuldis}
{\rm The assumptions that (1) the infectious period is non-random, (2) the infectious period starts immediately at infection, and (3) that the contacts are described by homogeneous Poisson processes, are too strong. They might be replaced by the assumption that for every (ordered) pair of individuals  the event that the first individual, if infectious, contacts the second individual is independent of contacts between other pairs of individuals.
In particular, the methods and results of this manuscript apply to models in which the infectious periods of individuals are not random. This inclused SEIR epidemic models with non-random infectious period, in which there is a random exposed (latent) period  between the moment an individual is infected and the moment that it starts to be infectious.  See \citep{Kuul82,Mees11} for a discussion.}
\end{remark}

\subsection{Results}
Our results are twofold. In the first place we introduce certain branching processes (sometimes multi-type) which  enable us to carefully describe the initial phase of an epidemic. As far as we are aware, our methods for computing the quantities of interest in the hierarchical model, in particular the approximating branching process, are new. The various branching processes used for the two models are somewhat hard to compare directly since the units of the various branching processes are not the same. In order to use the branching processes and make actual computations, we need to know how to make exact computations for epidemics restricted to households or schools, which are relatively small compared to the total population. This part is carried out in Section 3, while the actual branching process approximations are described in Section 4.
Our strategy for the independent partition model is similar to the computations suggested by \cite{Ball02} and our results are in agreement with theirs. We note that \cite{Ball02} do allow for random infectious period and their model is in that sense more general than ours. 

In the second place we are interested in direct comparison of the independent and the hierarchical model. We prove the following theorem.
\begin{theorem}\label{mainthm}
Consider a hierarchical and independent partition model in which households and schools have non-random sizes. Let $n$ be the number of schools and let $Z^H(n)$ and $Z^I(n)$ denote the number of ultimately recovered individuals in the hierarchical model and independent partition model, respectively.  Then, for any fixed $k$, we have
\begin{equation}\label{mainthmeq}
\liminf_{n\to \infty}\left\{ \mathbb{P}(Z^H(n)\leq k) - \mathbb{P}(Z^I(n)\leq k) \right\}\geq 0.
\end{equation}
\end{theorem}
Hence for fixed household and school sizes, the epidemic spreads easier in the independent model. Note that the theorem implies that the probability of a large outbreak in the hierarchical model is bounded above by the probability of a large outbreak in the independent partition model. 

The assumption that all households and all schools have non-random sizes cannot be deleted in general. This is shown in the following theorem. 

\begin{theorem}
\label{mainthm2}
If we allow for variation in the sizes of the household and school in the population, then (\ref{mainthmeq}) does not hold in general. In particular, we have the following two counterexamples.

Let $p_h:=1-e^{-\lambda_h}$ and $p_s:=1-e^{-\lambda_s}$. In either of the following two situations, (\ref{mainthmeq}) does not hold for $j$ large enough:
\begin{enumerate}
\item For some fixed (and large) $j$, households have size $j$ with probability $(2j)^{-1}$, and size 1 with probability $1-(2j)^{-1}$. Furthermore, $N_c \equiv 1$, that is, in the hierarchical model schools contain exactly one household. Furthermore, $p_s = p_h = 2(3j)^{-1}$ and $\lambda_g=1/10$.\\
\item All households have size 2, so $N_h \equiv 2$. For some fixed (and large) $j$, $N_c=j$ with probability $(4j)^{-1}$ and $N_c=1$ with probability $1-(4j)^{-1}$.
Furthermore, $\lambda_g=1/20$, $p_h=1$ and $p_s=(3j)^{-1}$.
\end{enumerate}
\end{theorem}

Of course one example would suffice to show that  (\ref{mainthmeq}) does not hold in general, but the reason why (\ref{mainthmeq}) does not hold is somewhat different in the two examples, and therefore we present them both.

We prove these theorems in Section \ref{bewijs}.

\begin{remark}
{\rm The examples in Theorem \ref{mainthm2} are obviously extreme and chosen such that we can exploit dependencies which do not appear if all households and schools have the same size. In the first example there are huge differences between the sizes of the households. Individuals which are part of a large household are automatically part of a large school in the hierarchical model, while the sizes of the school and the household of an individual are independent in the independent partition model. The dependence can be used to increase the offspring mean of the  branching process approximating the epidemic in the hierarchical model, defined in Section \ref{bpapp}. In particular, the branching process can become supercritical in the hierarchical model, while without the dependencies it would be sub-critical.

In the second example, all households have the same size and are relatively small, but the variance in school sizes is large. Here we use that household members of individuals in a large school are also part of a large school in the hierarchical model, while this is not automatically the case in the independent partition model. In particular, the parameters of the model are chosen in such a way that if we ignore all global contacts, expected epidemic sizes will be larger in the hierarchical model than in the independent partition model. 

Less extreme examples could be used, but computations would be messy and the examples would rely on the same principles.}
\end{remark}

\section{Random graphs and local epidemics}

It is useful to describe the collection of ultimately recovered individuals by means of a random graph in which vertices represent individuals. This is a classical approach, see e.g.\ \citep{Cox88}. The graph is built up as follows. For every vertex we draw, with probability $p_h=1-e^{-\lambda_h}$, directed ``household edges'' to each of the other vertices corresponding to individuals in its household. Similarly we draw, with probability $p_s=1-e^{-\lambda_s}$ directed ``school edges'' to each of the other vertices representing individuals in its school. Finally we draw, with probability $p_g=1-e^{-\lambda_g/(N-1)}$, directed ``global edges'' to each of the other vertices in the population. All edges are drawn independently of each other.

The endpoints of respectively household, school or global edges starting at a given vertex correspond to the individuals that will be contacted by the individual represented by this given vertex via respectively household, school or global contacts during its infectious period, were it to be infected in the course of the epidemic. The set of individuals infected in the course of the epidemic started at a randomly chosen individual is distributed as the set of vertices that can be reached by a directed path starting at the vertex representing such a randomly chosen individual.

It is well known \citep{Cox88} that the set of vertices which can be reached from the vertex representing the initially infected individual has the same distribution as the cluster of this vertex in an {\em undirected} graph in which undirected (household) edges are drawn independently between (pairs of) vertices representing members of the same household with probability $p_h$, undirected (school) edges are drawn independently between vertices representing members of the same school with probability $p_s$ and finally, all pairs of vertices share a (global) edges with probability $p_g$. In this undirected graph, the {\em epidemic generating graph}, there is no time evolution. The \textit{school cluster} of a vertex is the cluster of the vertex in the epidemic generating graph if all global and household edges are ignored. This cluster corresponds to the individuals infected through a school epidemic if the individual corresponding to the index vertex gets infected. The \textit{household cluster} of a vertex is the cluster of the vertex in the epidemic generating graph if all global and school edges are ignored. The household cluster has a similar interpretation as the school cluster.

We state without proof that for both the hierarchical and the independent model, the epidemic generating graph has, in the large population limit with probability tending to 1, at most one cluster of the same order of size as the population. The fraction of the number of vertices in this cluster converges in probability to a constant as the number of schools goes to infinity. The proof of this fact runs along the lines of the proofs of similar statements for Erd\H{o}s-R{\'e}nyi graphs \citep{Durr06} (cf.\ \citep{Ball09}).
It implies that if the initially infected individual  is chosen uniformly at random from the population, then the probability of a large outbreak is the same as the fraction of individuals which are ultimately removed. Indeed, if we first construct the epidemic generating graph and then choose the vertex representing the initial infectious individual uniformly at random, then it is straightforward to see that the probability of a large outbreak and the fraction of the individuals which are infected during such an outbreak are both equal to the fraction of the vertices in the large cluster.

Next we present exact computations in small populations. Since we apply this to either household or school epidemics, we consider the situation in which there is only one type of contact, that is, the epidemic generating graph is built up in such a way that undirected edges between any pair of vertices exist with probability $p$, independently of all other pairs. The result is a special case of \citep[p.281]{Diek12}. 
 
\begin{theorem}[\citep{Diek12}]\label{fsthm}
Consider a standard SIR epidemic in a homogeneously mixing population in which initially 1 individual is infectious and $L$ individuals are susceptible. Let, for $0 \leq k \leq L$, $P_k^L=P_k^L(p)$ denote the probability that $k$ members out of the initial population of $L$ susceptibles are ultimately recovered. Then we have for all $ 0 \leq \ell \leq L$,
\begin{equation}\label{singlefs2}
\sum_{k=0}^{\ell} P_k^{L} {L-k \choose \ell-k}(1-p)^{-(L-\ell)(k+1)}= {L \choose \ell}.
\end{equation}
\end{theorem}

Applying this result for $\ell=0,1,\ldots, L$ consecutively enables us to efficiently compute $P_0^L, P_1^L,\ldots, P_L^L$.

This theorem has a multi-type variant \citep{Ball86}, which we will use in the analysis of the hierarchical model. Let $k$ be a positive integer. For $\boldsymbol{\ell} := (\ell_1,\ldots, \ell_k)$ and $\mathbf{L} := (L_1,\ldots, L_k)$, $\boldsymbol{\ell} \leq \mathbf{L}$ is defined to mean $\ell_i \leq L_i$ for all $1 \leq i \leq k$. Furthermore, we write
$$
{\mathbf{L} \choose \boldsymbol{\ell}} = \prod_{i=1}^k {L_i \choose \ell_i} \qquad \mbox{and} \qquad \sum_{\mathbf{u=0}}^{\boldsymbol{\ell}} = \sum_{u_1=0}^{\ell_1} \cdots  \sum_{u_k=0}^{\ell_k}.
$$
Let $p_{ij}$ be the probability that a given type-$i$ individual contacts a given type-$j$ individual during its infectious period in case the type-$i$ individual is infected. We assume that initially there is one infective individual, which without loss of generality can be chosen to have type 1.

\begin{theorem}[\citep{Ball86}]
Consider a population subdivided into $k$ different types of sizes $\mathbf{L} = (L_1+1, L_2,\ldots, L_k)$, where 1 individual of type 1 is initially infectious and all other individuals are initially susceptible. Let $P_{\mathbf{u}}$ be the probability that the vector of numbers of ultimately recovered individuals (not including the initial infective) in the epidemic is equal to $\mathbf{u} = (u_1,\ldots, u_k)$. Then for each $0 \leq \boldsymbol{\ell} \leq \mathbf{L}$ we have
\begin{equation}\label{multfs}
\sum_{\mathbf{u}=0}^{\boldsymbol{\ell}} {\mathbf{L-u} \choose \boldsymbol{\ell-u}} P_{\boldsymbol{\ell}}/\prod_{i=1}^k \left(\prod_{j=1}^k (1-p_{ij})^{L_i - \ell_j}
\right)^{\mathbf{1}_{\{i=1\}}+ u_i}={\mathbf{L} \choose \boldsymbol{\ell}}.
\end{equation}
\end{theorem}

\section{Branching process approximations}
\label{bpapp}

For SIR epidemics in large homogeneously mixing populations, branching process approximations are reasonable since it is unlikely that during the early stages of the epidemic contacts of infectious individuals are made with non-susceptibles (see e.g.\ \citep{Diek12} or use birthday problem arguments). However, within schools or households, the epidemics take place in small groups, and a standard branching process approximation is no longer viable. In this section, we explain that despite the existence of local epidemics inside schools or households, a branching process approximation can still be carried out. We carry out the approximations separately for the two types of models. We reserve the word ``child" to refer to the offspring of a particle in the branching processes, {\em not} to a child in the actual population.

\subsection{The independent partition model}

In this subsection we show that we can carry out a branching process approximating in the independent partition model, where the units are individuals of the population and which will have three types. The three types of particles in the branching process correspond to individuals infected through either global, school or household contacts respectively. The number of global children in the branching process of each particle is Poisson distributed with mean $\mu_G$. Particles corresponding to individuals not infected through a school contact have a number of ``school-children'' distributed as the size of a school epidemic of a randomly chosen individual in the population. Similarly, particles corresponding to individuals not infected through a household contact have a number of ``household-children'' distributed as the size of a household epidemic of a randomly chosen individual in the population. Below follows a more detailed approach.

Imagine that we know the number and sizes of the schools and the households in the partition, but that we have not assigned individuals to the partitions yet. Let $n$ be the number of schools and $n'$ be the number of households, and number the schools from $1$ to $n$ and the households from $1$ to $n'$.

If we choose an individual uniformly at random from the population, then the probability that we choose an individual from a household of size $k$ is given by the {\em size biased} distribution $k\mathbb{P}(N_h =k)/\mathbb{E}(N_h)$. Indeed, it is $k$ times more likely to be in a given household of size $k$ than it is to be in a  given household of size $1$. A size biased variant of a random variable is decorated with a tilde, so $\mathbb{P}(\tilde{N}_h=k):= k\mathbb{P}(N_h =k)/\mathbb{E}(N_h)$. Similarly the school size of a uniformly at random chosen individual is distributed as $\tilde{N}_s$ and the number of households within the school of a uniformly at random chosen individual is distributed as $\tilde{N}_c$. 

We now create an i.i.d.\ sequence of randomly chosen schools by picking schools with replacement according to a size biased distribution. (There is a technical detail here which we ignore: in order to draw with replacement and to perform couplings below, we actually need to use the empirical distributions for household and school sizes, i.e.\ the distribution determined by the actual (finite) sequence of household and school sizes. A rigorous treatment of this detail can be found in \citep{Ball09}.) Say that this sequence is $x(1), x(2), \ldots$. Let $T_s$ be the first repeated index in this sequence, that is, \ $T_s = \min\{j:x(j)=x(i) \mbox{ for some $i<j$}\}$. Similarly create an infinite i.i.d.\ sequence of households drawn (with replacement) according to a size biased distribution. We denote this sequence by $x'(1), x'(2),\ldots$ and let $T_h$ be the first repeated index in the sequence of households. Since the school and household sizes have bounded support, a birthday type argument gives that for $c \in (0,1/2)$, we have $\mathbb{P}(T_s <n^c) \to 0$ and $\mathbb{P}(T_h <n^c) \to 0$ as $n\to \infty$. 

We now describe the coupled construction of the branching process and the epidemic generating graph. In the cluster of the initial infective in the epidemic generating graph and the approximating branching process we distinguish between three types of vertices: ``global", ``school" and ``household", where the type of a vertex is the type of the edge through which the vertex enters the process. The ancestor of the branching process (and the uniformly at random chosen vertex used to start the exploration of the epidemic generating graph) receives the type ``global".

We associate to the ancestor in the branching process school $x(1)$ and household $x'(1)$. Assume that school $x(1)$ has size $l$, then the number of children of the ancestor with type ``school" is equal to $k$ with probability $P_k^{l-1}(p_s)$. (Here, the $-1$ in $l-1$, comes from the fact that in this school epidemic there are $l-1$ initial susceptibles). If the household $x'(1)$ has size $l'$ then the number of children of the ancestor with type ``household" is equal to $k'$ with probability $P_{k'}^{l'-1}(p_h)$, and this number is independent of the number of ``school'' children. Finally, the ancestor also has a Poisson number of ``global" children with expectation $\lambda_g$. This number is independent of the number of ``household" and ``school" children. 

Important to note is that, in order to be able to use branching process approximations, we treat all vertices in a school/household cluster of the ancestor in the epidemic generating graph as its children, while in reality the ancestor and child might not share an edge, but only have a path of school/household edges between them. We need this technique of assigning vertices to a generation to keep branching process approximations meaningful in the sense that we retain enough independence (cf.\ \citep{Pell12}).

To the ``global'' children we assign both households and schools: if the number of ``global'' children is $k$ then we assign schools $x(2)$ up to and including $x(k+1)$ and households $x'(2)$ up to and including $x'(k'+1)$ to those children. Furthermore, we assign households to the ``school'' children (the following households in the sequence) and schools to the ``household'' children (the following schools in the sequence). As long as the total number of assigned schools is less than $T_s$ and the total number of assigned households is less than $T_h$, we can proceed with the construction in the obvious way. In this construction we create part of the epidemic generating graph through a branching process.

If the coupling proceeds then we assign vertices to school $x(1)$ and household $x'(1)$ in such a way that the household and school overlap at only one vertex, say vertex $v$, which is the vertex corresponding to the ancestor in the branching process. If in the branching process the ancestor has $k$ ``school'' children, then the size of the cluster of $v$ created by school edges is $k+1$ (the $+1$ is because $v$ is also part of the cluster). 
In the same way we choose the cluster of $v$ created by household contacts in $x'(1)$. If the number of ``household'' children of the ancestor is $k'$, then this cluster has size $k'+1$. Finally, global edges are drawn to vertices  which are part of households and schools which (i) only overlap with each other at the chosen vertices itself, and (ii) do not overlap with the households and schools of vertices in households and schools already explored.

The next step is to assign individuals (and the schools and households they belong to) to the generation 1 vertices. This happens in exactly the same way as individuals were assigned to the ancestor, apart from the fact that ``school'' individuals do not have ``school'' children, since their school is already explored,  and ``household'' individuals do not have ``household'' children. As long as the number of schools assigned to individuals does not exceed $T_s$ and the number of households assigned to individuals does not exceed $T_h$, the construction proceeds.

Since the probability that $T_s$ or $T_h$ is less than $n^{c}$ for $c \in (0,1/2)$ goes to 0 as $n \to \infty$, the branching process approximation works with large probability for all small clusters (i.e.\ clusters of smaller order than $n^{1/2}$)  in the epidemic generating graph. If the vertex representing the initial infective individual is in such a cluster then the approximating branching process goes extinct. If the initial infective individual is part of a large cluster, i.e.\ a cluster which asymptotically contains a positive fraction of the graph, then standard arguments used in random graph theory (e.g \citep[Ch.\ 3]{Durr06}) show that the approximating branching process survives.

The next step is to compute the probability that the constructed branching process dies out. Every individual has a Poisson number of ``global" children with mean $\lambda_g$. Every ``global" and ``school" individual has a random number of ``household" children, and we denote the probability that this number is equal to $k$ by 
$$
z_{h}(k) := \sum_{l=k}^{\infty}\mathbb{P}(\tilde{N}_h=l+1) P^l_k(p_h),
$$ 
where $P^l_k(p_h)$ is defined via (\ref{singlefs2}).
Similarly every ``global" and ``household" individual has a random number of ``school" children. We denote the probability that this number is equal to $k$ by 
$$
z_{s}(k) := \sum_{l=k}^{\infty}\mathbb{P}(\tilde{N}_s=l+1) P^l_k(p_s).
$$
``School" individuals do not have ``school" children and ``household" individuals do not have ``household" children, but apart from that the number of children of the different types are independent.

It is well known \citep[Ch.\ 4]{Jage75} how to compute extinction probabilities for multi-type branching processes. Define the probability generating functions of the offspring distribution as follows. For $0 \leq t_g, t_s, t_h \leq 1$, 
\begin{eqnarray*}
f_g(t_g,t_s,t_h) & := & \sum_{k_g=0}^{\infty}\sum_{k_s=0}^{\infty}\sum_{k_h=0}^{\infty}  \frac{(\lambda_g)^{k_g}}{k_g!}e^{-\lambda_g} z_s(k_s) z_h(k_h) (t_g)^{k_g} (t_s)^{k_s} (t_h)^{k_h},\\
\ & = & \sum_{k_s=0}^{\infty}\sum_{k_h=0}^{\infty}  e^{-\lambda_g(1-t_g)} z_s(k_s) z_h(k_h) (t_s)^{k_s} (t_h)^{k_h} 
\end{eqnarray*}
and similarly,
\begin{eqnarray*}
f_s(t_g,t_h) & := & \sum_{k_h=0}^{\infty} e^{-\lambda_g(1-t_g)} z_h(k_h)  (t_h)^{k_h}  \qquad (0 \leq t_g, t_h \leq 1),\\
f_h(t_g,t_s) & := & \sum_{k_s=0}^{\infty} e^{-\lambda_g(1-t_g)} z_s(k_s)  (t_s)^{k_s}  \qquad (0 \leq t_g, t_s \leq 1).
\end{eqnarray*}
Recall that we assume that the initial infective individual is a global individual. The probability of extinction of the branching process is equal to $t_g$, where $(t_g,t_s,t_h)$ is the smallest positive real solution of the following set of equations:
\begin{eqnarray*}
t_g & = & f_g(t_g,t_s,t_h),\\
t_s & = & f_s(t_g,t_h),\\
t_h & = & f_h(t_g,t_s).
\end{eqnarray*}

It is equally well known \citep[Ch.\ 4]{Jage75} how to quickly decide whether or not the probability that the approximating branching process survives is positive. Let $m_s = \sum_{k=0}^{\infty} k z_s(k)$ be the expected number of individuals infected in a school epidemic (excluding the initially infected individual in the school)  and $m_h = \sum_{k=0}^{\infty} k z_h(k)$ be the expected number of individuals infected in a household epidemic (exuding the initially infected individual in the household). Define the so called next generation matrix of the branching process by
\begin{equation}
M= \left(
\begin{array}{ccc}
\lambda_g & m_s & m_h\\
\lambda_g & 0 & m_h\\
\lambda_g & m_s & 0
\end{array}
\right).
\end{equation}
The probability of extinction of the branching process is strictly less than 1 if and only if the largest eigenvalue of $M$ (which is positive and real) is strictly larger than 1. A small computation shows that this is the case exactly when 
\begin{equation}
\label{matrix}
\lambda_g (m_s +1)(m_h+1) > 1-m_sm_h = (m_s +1) + (m_h +1) -  (m_s +1)(m_h+1).
\end{equation}
These results corresponds to the results in \citep{Ball02}. This largest eigenvalue of $M$ is often referred to as the basic reproduction number, $R_0$ \citep{Diek12,Pell12}. 
In Section \ref{bewijs} we will give some examples where (\ref{matrix}) is used.

\subsection{The hierarchical model}

In the hierarchical model the spread within its household and within its school caused by an individual are no longer independent, since households are entirely contained in schools. We get around this problem by changing the unit of the branching process. In particular, the particles in the branching process no longer correspond to individuals, but to clusters of individuals in the same household. Moreover, the type of a particle is given  by the number of individuals in its corresponding cluster. Below we derive the offspring distribution for this branching process. However, note that there are no easy closed expressions available for describing this distribution. 

Consider a household of size $k$, say. The epidemic generating graph restricted to this household and restricted to household edges, partitions the household into clusters. The joint distribution of the sizes of the clusters in the partition can be obtained via (\ref{singlefs2}). Indeed, the size of the first cluster (in order of exploration) is $l_1$ with probability $P^{k-1}_{l_1-1}$. Conditional on the size of the first cluster being $l_1$, then the size of the second cluster is $l_2$ with probability $P^{k-l_1-1}_{l_2-1}$, and so on. In this way, every household is partitioned into clusters with the property that if one of the individuals in the cluster gets infected, then all the vertices in that cluster get infected. Further infections within that household have to go through either school and global contacts, or through individuals outside the household.

Instead of considering a school as partitioned into households, we view a school as partitioned into clusters generated by household edges. The sizes of those clusters are not independent. The joint distribution of these cluster sizes is difficult to describe explicitly, but, as described in the previous paragraph, computationally relatively easy to deal with.

To compute the final size of an epidemic restricted to  school and household contacts for a school with a given configuration of households in it, we first assign types to the clusters generated by household contacts, where the type is the number of vertices within the cluster. Those clusters are the ``super-individuals'' and we apply (\ref{multfs}) to compute the final size within the school, with $p_{ij} = 1-(1-p_s)^{ij}$. Note that $1-p_{ij}$ is the probability that there is no school contact between any of the individuals in the type $i$ cluster and any of the individuals in the type $j$ cluster. In order to compute the number of individuals infected in an epidemic restricted to a school, let $u_i$ denote the number of clusters of size $i$ which are ultimately infected through school contacts (now including the initially infected cluster within the school), which can be computed by using (\ref{multfs}). The final size of the epidemic restricted to household and school contacts is then $\sum_{i=0}^{\infty}i u_i -1$. (The $-1$ originates from the fact that the initial infective individual within a school is not included in its final size.) Even for moderately large household and school sizes this sum has already many terms and we refrain from making the sum explicit. 

In the previous paragraph we assumed that the sizes of the households in a school are known. In order to perform further computations, we need to describe what the distribution is of such configurations for schools which are affected by the epidemic during the early stages of the epidemic.
To do this we observe that if we choose a vertex uniformly at random then its household size is size biased and distributed as $\tilde{N}_h$. The number of households in the school this vertex is part of is then distributed as $\tilde{N}_c$. All other households in this school have sizes distributed as $N_h$. We can use this to find the distribution of the size of the cluster created by school and household edges which contains a uniformly at random chosen vertex from the population. Let $Y$ have the same distribution as this random variable. Note that the uniformly chosen vertex is incorporated in $Y$. The distribution of $Y$ is difficult to describe in closed form, but it is computationally tractable.

Finally we can describe the branching process. Since school sizes have bounded support, a giant component in the epidemic generating graph means that many schools are infected. This suggests that we can consider a branching process of {\em initial} cases in schools (and therefore a branching process of infected schools cf.\ \citep{Ball02}). The (direct) offspring of a particle of the branching process consists of all vertices that can be reached by a path of school and household edges, apart from the final edge which is global and leads to a new infected school. So, the direct offspring of a particle correspond to all vertices which can be reached by a global edge from one of the vertices corresponding to an epidemic restricted to household and school contacts from the individual corresponding to the particle. If this approximating branching process survives then 
the vertex corresponding to the initial infectious individual is in the giant component
(with large probability as $n \to \infty$), while if this branching process goes extinct than the cluster of the vertex corresponding with the initial infectious individual is also small (compared to $n$).
The number of children of an individual in the branching process is distributed as $Z \sim \sum_{k=1}^Y X_k$, where the $X_k$'s  are i.i.d.\ Poisson random variables with mean $\lambda_g$, which are independent of $Y$. This branching process is a single type branching process for which the extinction probability, $q$, is the smallest root of  $t = \sum_{k=0}^{\infty} \mathbb{P}(Z=k)t^k$ \citep{Jage75}.
This smallest root is strictly less than 1 if and only if the offspring mean $R_* = \mathbb{E}(Y)\lambda_g >1$. We give some examples of how to use these computations in the proof of Theorem  \ref{mainthm2} below.

\section{Proof of Theorems \ref{mainthm} and \ref{mainthm2}}
\label{bewijs}

\textit{Proof of Theorem \ref{mainthm}.}
We use a coupling of the epidemic processes on the hierarchical and the independent partition models similar to the one used for the branching process approximations in the previous section. This coupling is then used to show that $Z^I(n)$ and $Z^H(n)$ are asymptotically stochastically ordered as stated in Theorem \ref{mainthm}.

Again we consider an i.i.d.\ sequence $x(1), x(2), \ldots$ of schools, by uniformly picking schools with replacement. Let $T_s= T_s(n)$ be the index of the first repeated school in this sequence. Similarly create an infinite i.i.d.\ sequence $x'(1), x'(2),\ldots$ of households drawn uniformly with replacement, and let $T_h=T_h(n)$ be the index of the first repeated household in this sequence. Let $\mathcal{A}_k = \mathcal{A}_k(n) =\{T_s >k+1\} \cap \{T_h>k+1\}$.
We have
$$
\mathbb{P}(Z^I(n) \leq k) \leq \mathbb{P}(\{Z^I(n) \leq k\} \cap \mathcal{A}_k) + \mathbb{P}(\mathcal{A}_k^c). 
$$
By birthday-problem type arguments we have that for all $\delta>0$,  $\mathbb{P}(\mathcal{A}_k^c)< \delta$ for sufficiently large $n$. Since $\mathbb{P}(\{Z^H(n) \leq k\} \cap \mathcal{A}_k) \leq \mathbb{P}(Z^H(n) \leq k)$, it suffices to prove that 
\begin{equation}\label{help1}
\mathbb{P}(\{Z^I(n) \leq k\} \cap \mathcal{A}_k)  \leq \mathbb{P}(\{Z^H(n) \leq k\} \cap \mathcal{A}_k),
\end{equation} 
or equivalently, 
\begin{equation}\label{help2}
\mathbb{P}(Z^I(n) \leq k|\mathcal{A}_k)  \leq \mathbb{P}(Z^H(n) \leq k|\mathcal{A}_k).
\end{equation} 

We simultaneously construct the epidemic generating graph of the hierarchical and of the independent partition model on a suitable probability space and show that for all $k$, and conditioned on $\mathcal{A}_k$, the inclusion $\{Z^{I}\leq k\} \subseteq \{Z^H \leq k\}$ holds.

Let $x(1)$ be the school of the initially infected individual both for the hierarchical and the independent partition model. Let a school epidemic run in this school and use this epidemic for both the hierarchical and the  independent partition model. Note that this gives the right distribution of the first school epidemic in both models. 

Now assume that the size of the school cluster in the corresponding epidemic generating graph is $j$, where  $j \leq k$. In the independent partition model the households of the $j$ individuals already infected are $x'(1), \ldots, x'(j)$. Those households do not overlap since we condition on $\mathcal{A}_k$. The size of the household epidemics (including the initial infected within the household) are then i.i.d.\ and all distributed as the random variable $X$, where $\mathbb{P}(X=i) = P^{n_h-1}_{i-1}$, where $P^{n_h-1}_{i-1}$ is defined as in Theorem \ref{fsthm}.  In the hierarchical model, the households of the $j$ individuals affected by the school epidemic do not need to  be all different. The household epidemics are now run one by one, the initial infectives for those household epidemics are the $j$ individuals affected by the school epidemic, we do however ignore the individuals already infected before (by the school epidemic, or by household epidemics explored earlier). The probability that $i$ individuals are ultimately infected through such a household epidemics is given by  $P^{\tilde{n}_h}_{i-1}$, where $\tilde{n}_h$ is the number of individuals in the household not affected before by the epidemic. Observing that  $\tilde{n}_h \leq n_h-1$ gives that the number of vertices affected by such a local epidemic is always smaller than the same quantity in the independent partition model.

The next step is to investigate the school epidemics of the individuals infected through household contacts. Note that in the independent partition model, the number of susceptible schoolmates of the individual infected through a household contact is $n_s-1$ (by the conditioning on $\mathcal{A}_k$), while in the hierarchical model this number is at most $n_s-2$. We proceed in this way analysing the epidemic through school and household contacts and we notice that the size of the cluster generated by school and household edges in the epidemic generating graph in the hierarchical model is bounded above by the same quantity as in the independent partition model. Let $C_1^H$ (respectively $C_1^I$) denote this cluster in the hierarchical model (respectively independent partition model). If the number of vertices in such a cluster in the hierarchical or independent partition model is at most $l$, then the number of households and schools investigated is at most $l$ since each vertex is part of exactly 1 school and 1 household.

The next step is to investigate the global edges from the vertices in $C_1^H$ and $C_1^I$, assume that $C_1^H$ has size $l'$  and $C_1^I$ has size $l$.  Note that $l'\leq l$. We keep the two epidemic generating graphs coupled, by using a sequence of $l'$ i.i.d.\ Poisson numbers with mean $\lambda_g$ and use these for the number of global contacts of the first $l'$ vertices in $C_1^I$ and for the $l'$ vertices in $C_1^H$. In addition we independently assign i.i.d.\ Poisson numbers (with mean $\lambda_g$) to the other $l-l'$ vertices in  $C_1^I$. Note that if $l$ plus the total number of global edges from $C_1^I$ does not 
exceed $k$, then,  because of the conditioning on $\mathcal{A}_k$, the globally contacted vertices are all in different households and schools and also not in households and schools encountered before. In the hierarchical model we assign the same schools to the globally contacted vertices as in the independent partition model or a subset of those. Then we proceed with investigating the epidemic generating graph by investigating the new schools in the same way as we investigated the school of the initially infected vertex.

By this coupling we obtain that if the exploration of the cluster stops before $k\!+\!1$ vertices are included in the independent partition model, then in the hierarchical model the cluster size is also less than $k\!+\!1$. Note that if we have explored the cluster in the independent partition model until we have included $k$ vertices and there are still school, household or global edges not yet explored in the construction then conditioning on $\mathcal{A}_k$ guarantees  that this $k\!+\!1$-st edge is to a not yet encountered vertex and so the cluster of the initially chosen vertex is at least $k\!+\!1$. This guarantees that conditioned on $\mathcal{A}_k$, we have   
$\{Z^{I}\leq k\} \subseteq \{Z^H \leq k\}$ and completes the proof. 

\medskip\noindent
\textit{Proof of Theorem \ref{mainthm2}.}
We start the computations with Example 1. In this example households have size $j$ with probability $(2j)^{-1}$, and size 1 with probability $1-(2j)^{-1}$. Furthermore, in the hierarchical model schools contain exactly one household. We set $p_s = p_h = 2(3j)^{-1}$ and $\lambda_g=1/10$. We then choose $j$ large enough to support our claim. Observe that the fraction of individuals within a household of size 1 is given by
$$\frac{1-(2j)^{-1}}{1-(2j)^{-1} + j(2j)^{-1}}= \frac{2j-1}{3j-1} \approx \frac{2}{3}.$$

Consider the independent model. Since $j$ is large, we may approximate the epidemic within a household or school of size $j$ by a sub-critical branching process with offspring mean $2/3$. The total number of ultimately removed individuals in a school/household is then roughly $\sum_{k=0}^{\infty} (2/3)^k = 3$. In a school or household of size $1$, there are no secondary individuals. It is easy to check with (\ref{matrix}) that $M$ has largest eigenvalue less than 1, and therefore that the epidemic in the independent partition model is sub-critical.

In the hierarchical model, the probability of having an edge between two vertices in the same household is $1-(1-p_s)(1-p_h)(1-p_g)$ and since $p_s$, $p_h$ are small and $p_g$ is even much smaller than that ($p_g \approx \lambda/N$, where $N$ is the total population size), we have that $1-(1-p_s)(1-p_h)(1-p_g) \approx p_s +p_h =4(3j)^{-1}$. If $j$ is large this leads to a 
supercritical  Erd\H{o}s-R{\'e}nyi graph, if the epidemic generating graph is restricted to a combined household and school. 
In particular the largest cluster in a large school is of order $j$; say that it is with probability $1/2$ at least $\alpha j$, where $\alpha >0$. We claim that each large component is in expectation, via global edges, connected to $\lambda_g \times \alpha j \times 1/3 \times 1/2 \times \alpha $ 
other components of size at least $\alpha j$. Indeed, approximately one third of those edges connects to other vertices in large schools/households, half of them contains a combined household/school cluster of at least size $\alpha j$ and $\alpha$ is the probability that this edge actually ends in the large household/school cluster. If $j$ is large enough, the combined quantity is larger than 1 and there is a cluster of household/school clusters of size at least $\alpha j$  all connected through global edges which itself has size of order $N$.  This shows that the epidemic generating graph has a cluster of the same order of magnitude as the population, which implies that (\ref{mainthmeq}) does not hold. 

Next we consider Example 2. Recall that in this example all households have size 2. In the hierarchical model schools contain $j$ households with probability $(4j)^{-1}$ and only 1 household with probability $1-(4j)^{-1}$.
Furthermore, $\lambda_g=1/20$, $p_h=1$ and $p_s=(3j)^{-1}$.
Similarly to Case 1, we deduce that $m_s \approx 7/5$ and $m_h=2$. This implies, using (\ref{matrix}) again, that the epidemic in the independent partition model is subcritical.

In the hierarchical model, we see the households as ``super-individuals" (since if one of the two household members gets infected the other will automatically get infected as well). So we can consider a social structure which only contains schools. Two (super) individuals (i.e. households) contact each other with probability $1-(1-(3j)^{-1})^4 \approx 4(3j)^{-1}$. Again, the epidemic generating graph for the school epidemic is an Erd\H{o}s-R{\'e}nyi graph, in which vertices on average share edges with $4/3$ other vertices. Therefore, the Erd\H{o}s-R{\'e}nyi graph contains, with large probability, a cluster of order $j$ and we can copy the argument from the previous example.

\section{Discussion}

\begin{figure}[hp]
\centering
\subfloat[]{\label{fig:1}\includegraphics[width=0.45\textwidth]{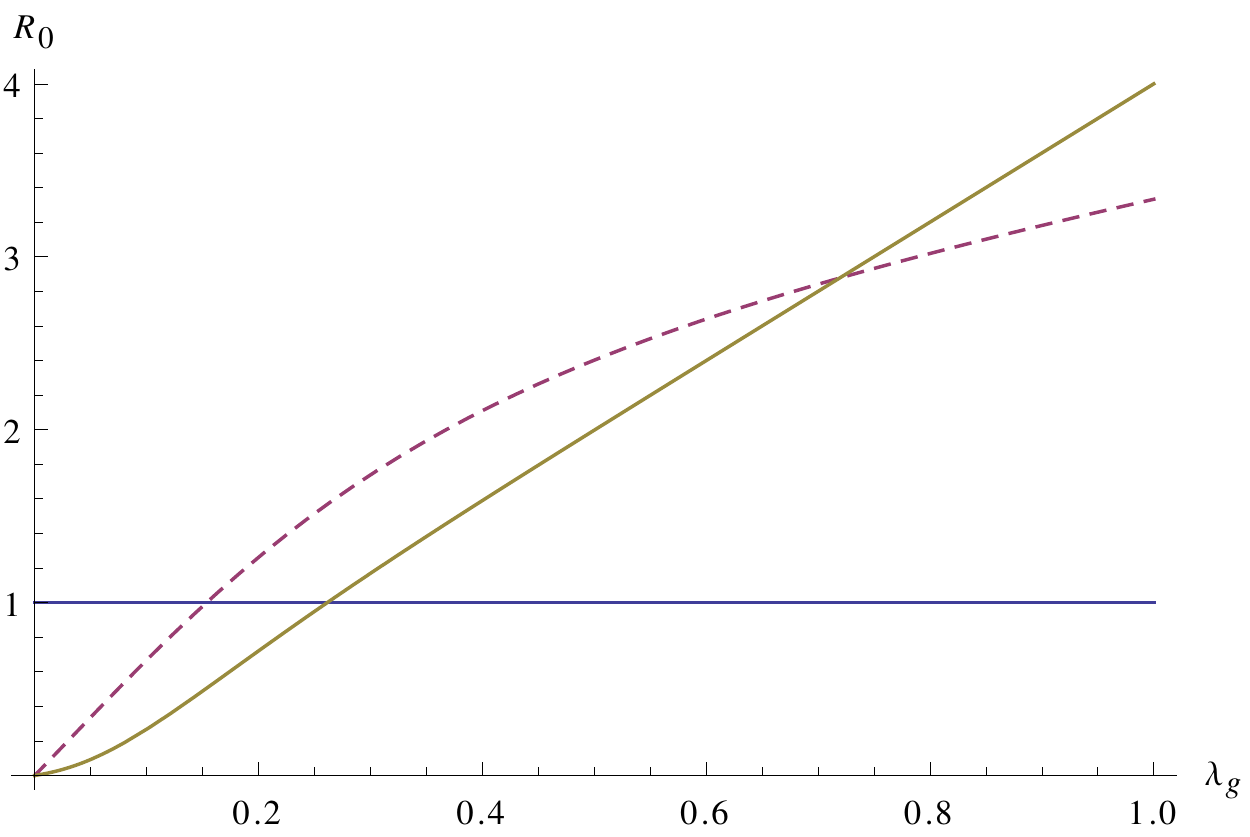}}
\subfloat[]{\label{fig:2}\includegraphics[width=0.45\textwidth]{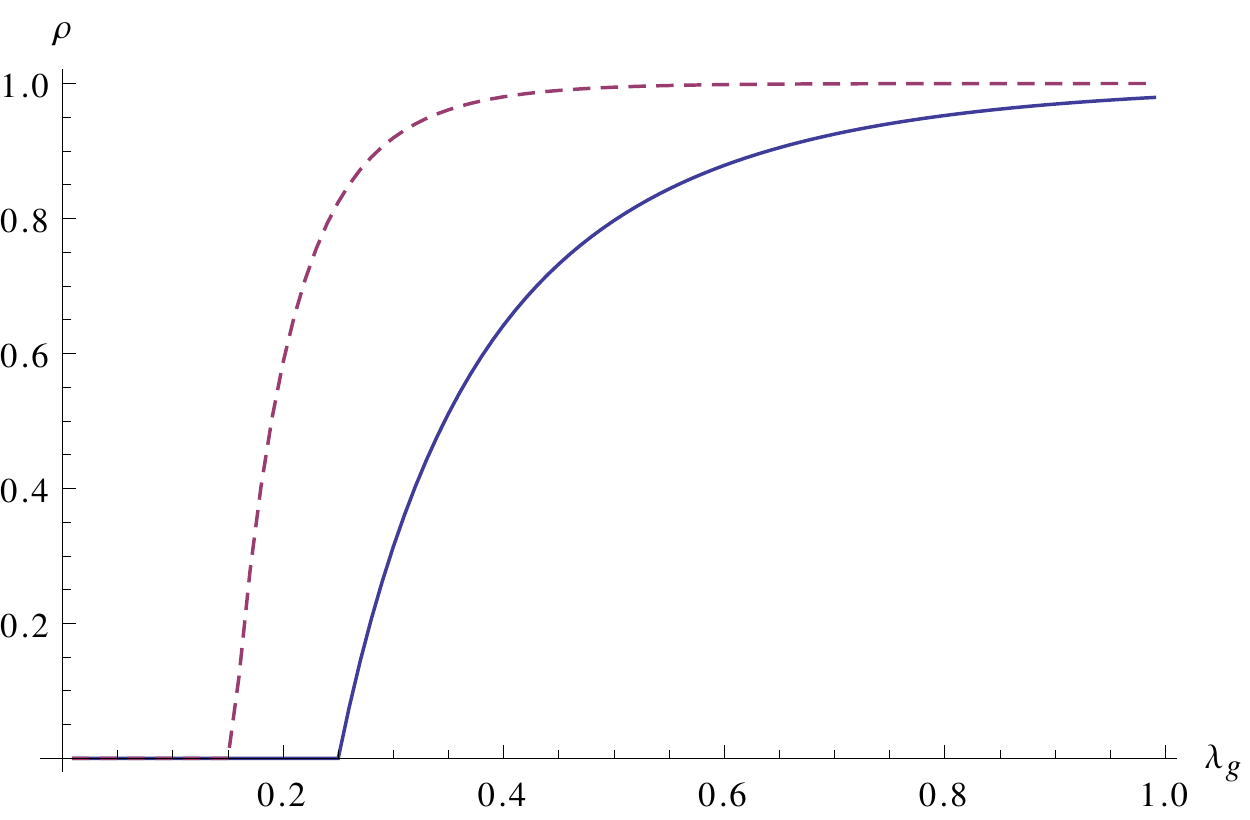}}
\caption{The basic reproduction number $R_0$ (a) and the survival probability $\rho$ (b) as a function of the global infection rate $\lambda_g$ for a model in which the proportions of the rates are $\lambda_g: \lambda_s : \lambda_h = 1:2:4$ and all households have size 2 and all schools have size 4. The hierarchical model is represented by the solid lines, while the independent partition model is represented by the dashed lines.}
\label{figure}
\end{figure}

We have discussed an ordering of epidemic severity of infectious diseases in two extreme population models. 
We stress that the comparisons are about the final size and the probability of a large outbreak, and not about the reproduction number $R_0$ (the offspring mean of the approximating branching process). Indeed, since the units of the branching processes are so different we have different interpretations of $R_0$ in the various models and it makes no immediate sense to compare this value for the two models directly. This is further illustrated in Figure \ref{figure} where we plot $R_0$ and the survival probability $\rho$ against the infection rates. In this figure we keep the proportions $\lambda_g: \lambda_s : \lambda_h$ fixed at $1:2:4$, while household sizes are 2 and schools have size 4. We note that while the survival probability of the independent partition model is at least as large as the survival probability for the hierarchical model, $R_0$ is not ordered in this way. However the infection rates for which $R_0$ crosses the threshold $1$, are lower for the independent partition model, as it should be, since the survival probability (for which there is an ordering) is strictly positive if and only if $R_0>1$. 
\bibliographystyle{abbrvnat}

\end{document}